\documentclass[onecolumn,showpacs,showkeys,amssymb,nobibnotes,superscriptaddress,aps,pra,12pt,preprint]{revtex4-1}
\usepackage{graphicx}
\usepackage{bm}

\begin{document}

\preprint{APS/123-diode}

\title{Nonreciprocity light propagation in coupled microcavities system beyond weak-excitation approximation}
\author{Anshou Zheng}\email{zaszas1\_1@126.com}
\author{Guangyong Zhang}
\author{Hongyun Chen}\author{Tingting Mei}
\affiliation{School of Mathematics and Physics, China University of Geosciences, Wuhan 430074, People's Republic of China}
\author{Jibing Liu}\email{liu0328@foxmail.com}
\affiliation{College of Physics and Electronic Science, Hubei Normal University, Huangshi 435002, People's
Republic of China}

\begin{abstract}
We propose an alternative scheme for nonreciprocal light
propagation in two coupled cavities system, in which a two-level quantum emitter is coupled to one of the optical microcavities.
For the case of parity-time (\textrm{PT}) symmetric system (i.e., active-passive coupled cavities system), the cavity gain can significantly
enhance the optical nonlinearity induced by the interaction between a quantum emitter and cavity field
beyond weak-excitation approximation. The giant optical nonlinearity results in the non-lossy nonreciprocal light
propagation with high isolation ratio in proper parameters range. In addition, our calculations show that nonreciprocal
light propagation will not be affected by the unstable output field intensity caused by optical bistability and we can even switch
directions of nonreciprocal light propagation by appropriately adjusting the system parameters.

\end{abstract}


\maketitle

\section{Introduction}
Achieving rapid development in integrated photonic circuits depends on the all-optical elements, which
are essential for high-speed processing of light signals. Nonreciprocal light propagation is
an indispensable common trait for some optical elements, such as optical diodes, optical isolator,
circulator, etc. For example, the optical diode permits the light transport in only one direction but not
the opposite direction. The successful design of nonreciprocal light propagation devices relies
on the breaking of the time-reversal symmetry. Thus, nonreciprocal light propagation is inherently
difficult, even in theory because of the time-reversal symmetry of light-matter interaction \cite{1book}.
Motivated by the tremendous application of nonreciprocal electrical current propagation, an immense attention has been paid
to the study of nonreciprocal light propagation. As a traditional method, a material with strong magneto-optical
effects (faraday rotation) is often used to break the time-reversal symmetry for some optical devices \cite{2josa,3josa,4nat}.
However, unfortunately the requirement of the magneto-optical effect is the big size components and strong external magnetic fields,
which are harmful for the on-chip optical nonreciprocal devices. Beyond that, one can also break the time-reversal symmetry
and design the nonreciprocal optical devices by time-dependent effects \cite{5nat,6prl}, unbalanced quantum coupling \cite{j1,j2,j4,j9}
 or
optical nonlinearity \cite{7josa,8apl,j5,j6,9oe,10elec,11ol,j3}.
The ubiquitous optical nonlinearity in different optical systems has been extensively studied and further
adopted in design of nonreciprocal light propagation devices. For example, many schemes have been reported
through the nonlinearity of the waveguides, such as the second order nonlinearity $\chi^{(2)}$ \cite{7josa,8apl,j5,j6},
dispersion-engineered chalcogenide \cite{9oe}, Raman amplification \cite{10elec,11ol} and so on.

On the other hand, duce to the high-quality factor $Q$ and small mode volume $V$ of optical microcavities
\cite{12nature,13book,14science,15ar},
 it has attracted considerable interest for implementing
nonreciprocal light propagation devices \cite{16science,17prl,18oe,19apl,j7}. For instance, Fan et al. achieved the experiment of nonreciprocal
light propagation with the Kerr and thermal nonlinearity in silicon microring resonators \cite{16science}.
Based on a nonlinearity of an optomechanical system, some schemes of nonreciprocal behavior have also been reported
\cite{17prl,18oe,19apl,j7}. The above schemes, however, rely heavily on the strong nonlinearity, which is not easy to
obtain, especially for few-photon situations. Recently, some works show that the nonlinearity in the coupled
resonators can be greatly enhanced by the introducing optical gain in one resonator of the \textrm{PT}-symmetric system \cite{20pra,21prb}.
And an immense attention has been attracted to \textrm{PT}-symmetric system which has an interesting feature that
non-Hermitian Hamiltonian can still have an entirely real spectrum with respect to the \textrm{PT}-symmetry \cite{22prl,23pro}.
In addition, two coupled resonators can be processed as a \textrm{PT}-systemic system
\cite{24nat,25nat,26science,27science,28science,29prl}.
More recently, a few of schemes of nonreciprocal light propagation have been proposed with \textrm{PT}-systemic coupled resonators system
\cite{21prb,24nat,25nat}. For example, based on the inherent nonlinearity (i.e., gain-induced nonlinearity)
of the \textrm{PT}-systemic system, successful experiment has been carried out for nonreciprocal light propagation
with two coupled whispering-gallery-mode (WGM) microresonators \cite{24nat,25nat}. Note that through mechanical Kerr nonlinearity,
a theory scheme is also proposed for nonreciprocal phonon propagation with coupled mechanical resonators \cite{21prb}.
The weak mechanical Kerr nonlinearity is greatly improved by the gain in one mechanical resonator of the
\textrm{PT}-symmetry and results in the non-lossy and high isolation ratio nonreciprocal phonon propagation.

In this paper, we explore the optical nonlinearity induced by a single quantum emitter coupled to a microcavity beyond
weak-excitation approximation. Based on the optical nonlinearity, an alternative scheme is proposed for nonreciprocal
light propagation in a system of two coupled cavities and a single quantum emitter coupled to one of the cavities.
The scheme reported here has some important features. (i) The optical nonlinearity of the hybrid system is greatly
enhanced by the cavity gain. And the giant nonlinearity leads to the non-lossy nonreciprocal light propagation
with extremely high isolation ratio. (ii) Nonreciprocal light propagation means allowing transport of light in one
direction but not the opposite direction \cite{30science}. However, through adjusting proper parameters, to switch
between the blocking and allowing directions is allowed for the present scheme.
For different nonreciprocal light propagation cases,
we can all obtain the non-lossy transmission with high isolation ratio in allowing directions.
In addition, it is achievable to switch between unidirectional and bidirectional transport regime. (iii) Optical bistability or
even optical multistability behavior is often induced by optical nonlinearity, and it will lead to unstable output.
When the disturbance and perturbation of the system parameters are strong enough, the output field intensity will switch between
the different metastable values of the optical bistability. Obviously, it is harmful for the nonreciprocal light propagation.
However, via choosing proper parameters range, we can avoid the interference of unstable output and obtain certain output intensity
even for the strong disturbance of parameters.

This paper is organized as follows: In Sec.\,II, the physical model considered and the
corresponding Hamiltonian of the system is introduced. By applying the Heisenberg-Langevin formalism,
we can obtain the evolution equations of the system. In Sec.\,III, we investigate carefully the evolution equations and get the
numerical steady solution. Then, we explore the spectra of the optical output fields and analyze the influence of parameters on
nonreciprocal light propagation.
Finally, we give a conclusion in Sec.\,IV.

\section{Physical model and basic formula}
We consider the setup as shown in Fig. 1, where two single-mode optical microcavities of frequencies $\omega_{1(2)}$ are directly
coupled to each other with strength $J$, which is very sensitive to the distance between two cavities.
 The optical cavities are denoted by bosonic annihilation and creation operators
$\hat{a}_j$ and $\hat{a}^\dagger_j(j=1,2)$, respectively.
A two-level quantum emitter with transition frequency
$\omega_e$ is embedded in the first (j=1) cavity, and the cavity mode $\hat{a}_1(\hat{a}_1)^\dagger$ is coupled to the quantum emitter
transition $|e\rangle\Longleftrightarrow|g\rangle$ with the coupling strength $g$. We take the input probe field
as $S_{in}=\varepsilon_pe^{-i\omega_pt}$, where $\omega_p$ and $\varepsilon_p$ is the carrier frequency and
the amplitude of the probe field propagating in the waveguide. The amplitude of the input probe field $\varepsilon_p$
is normalized to a photon flux at the input of the cavity and the directly related power is $P=\hbar\omega_p\varepsilon_p^2$.
Under the rotating-wave and the electric-dipole approximation, the effective Hamiltonian of the hybrid
\begin{figure}[htb]
\centerline{\includegraphics[width=.2\textwidth,height=0.45\textheight,angle=90,]{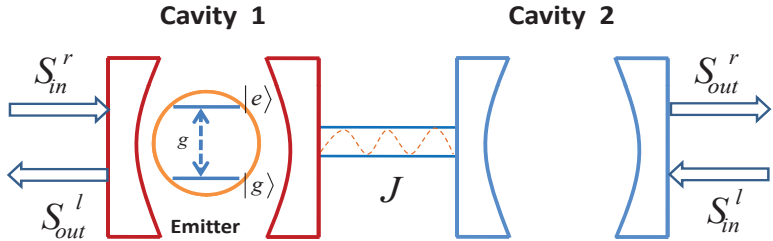}}
\caption{\label{fig1} (Color online) Schematic diagram of the hybrid optical system consisting of
 two coupled cavities and a two-level quantum emitter coupled to one of them.}
\end{figure}
optical system is written in the rotating frame at the frequency of the probe field $\omega_p$ as \cite{wu1,wu2}
\begin{eqnarray}\label{E1}
\hat{H}_e&=&\hbar(\Delta_1+\Delta_2)\hat{\sigma}_{ee}+\hbar\Delta_1\hat{a}_1^\dagger\hat{a}_1+\hbar\Delta_1\hat{a}_2^\dagger\hat{a}_2
+i\hbar g(\hat{a}_1\hat{\sigma}_{eg}-\hat{a}_1^\dagger\hat{\sigma}_{ge})\\\nonumber
&&+\hbar J(\hat{a}_1^\dagger\hat{a}_2+\hat{a}_2^\dagger\hat{a}_1)
+i\hbar\sqrt{\kappa_e}(\varepsilon_p\hat{a}_1^\dagger-\varepsilon_p^*\hat{a}_1),
\end{eqnarray}
where the symbols $\hat{\sigma}_{eg}(\hat{\sigma}_{ge})$ and $\hat{\sigma}_{ee}$ mean the transition operator and
 the population operator of the two-level quantum emitter, respectively.
We assume the two cavities have the same resonant frequencies $\omega_1=\omega_2=\omega_c$.
 $\Delta_1=\omega_c-\omega_p$ and $\Delta_2=\omega_e-\omega_c$  are the corresponding frequencies detunings of
 the cavity field frequency from the probe field frequency and the quantum emitter transition frequency, respectively.
The coupling parameter $\kappa_e$ describes the coupling loss rate between each cavity and the corresponding taper waveguide.
\begin{figure}[htb]
\includegraphics[width=0.4\textwidth]{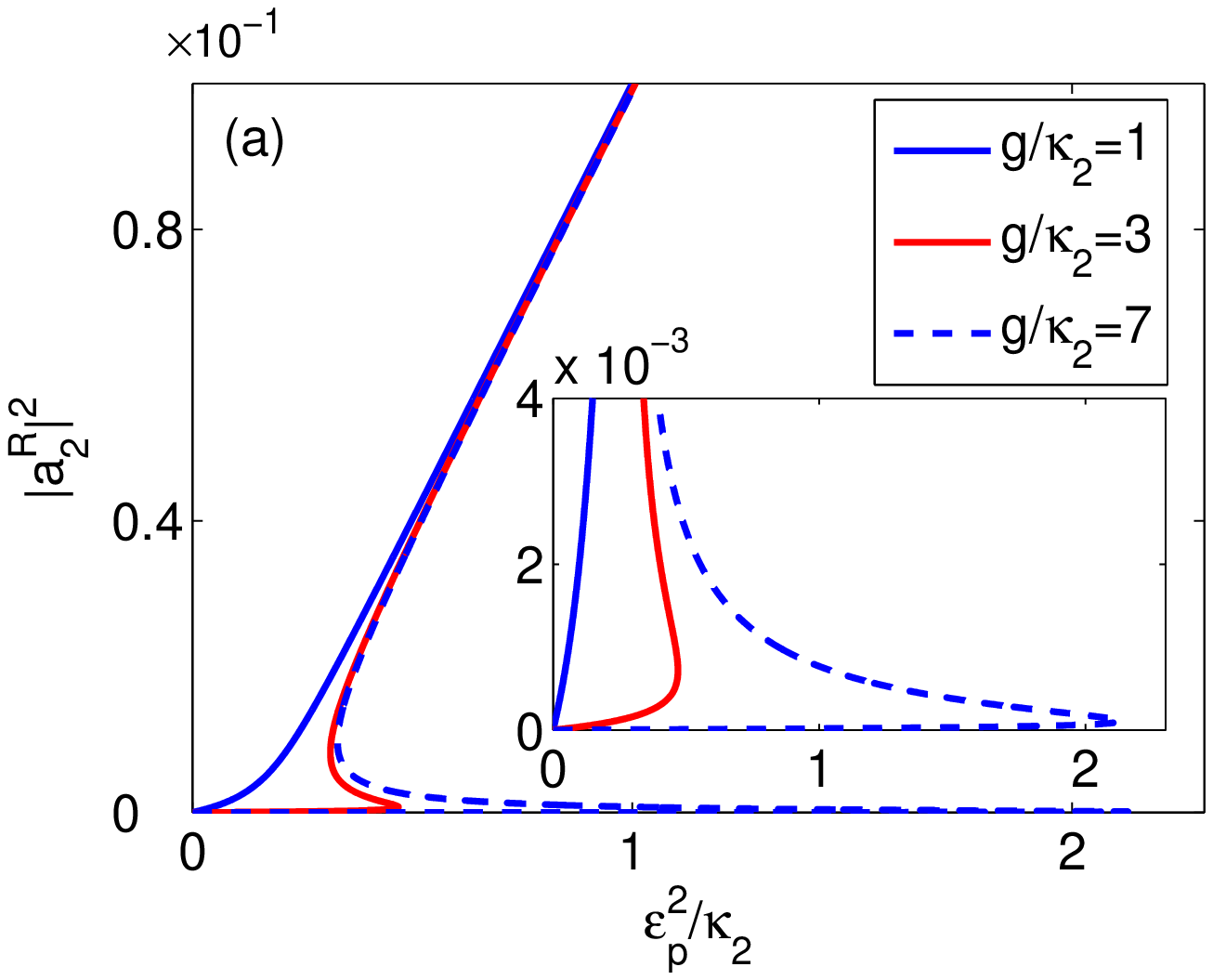}
\includegraphics[width=0.4\textwidth,height=0.2\textheight]{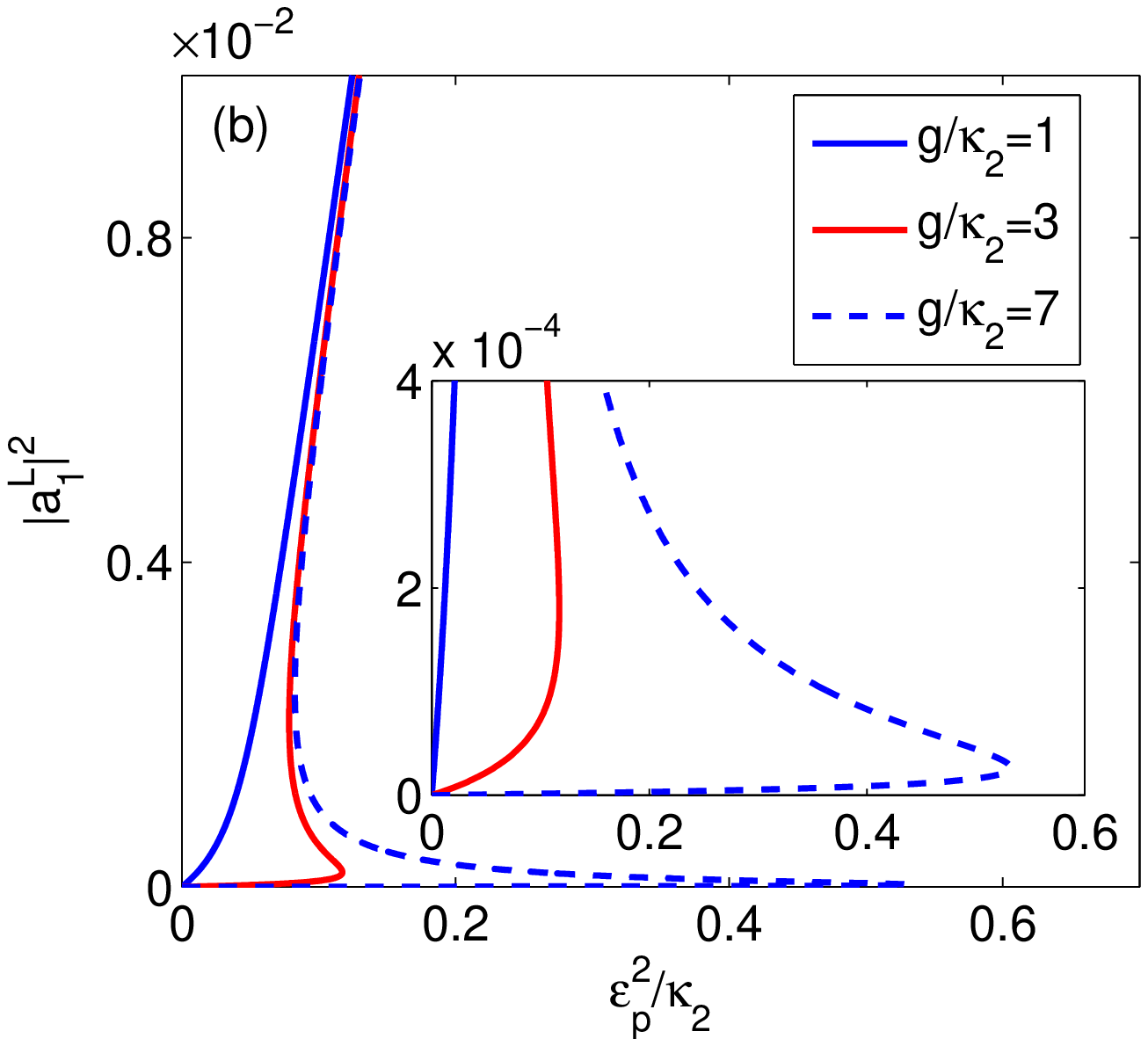}
\includegraphics[width=0.4\textwidth]{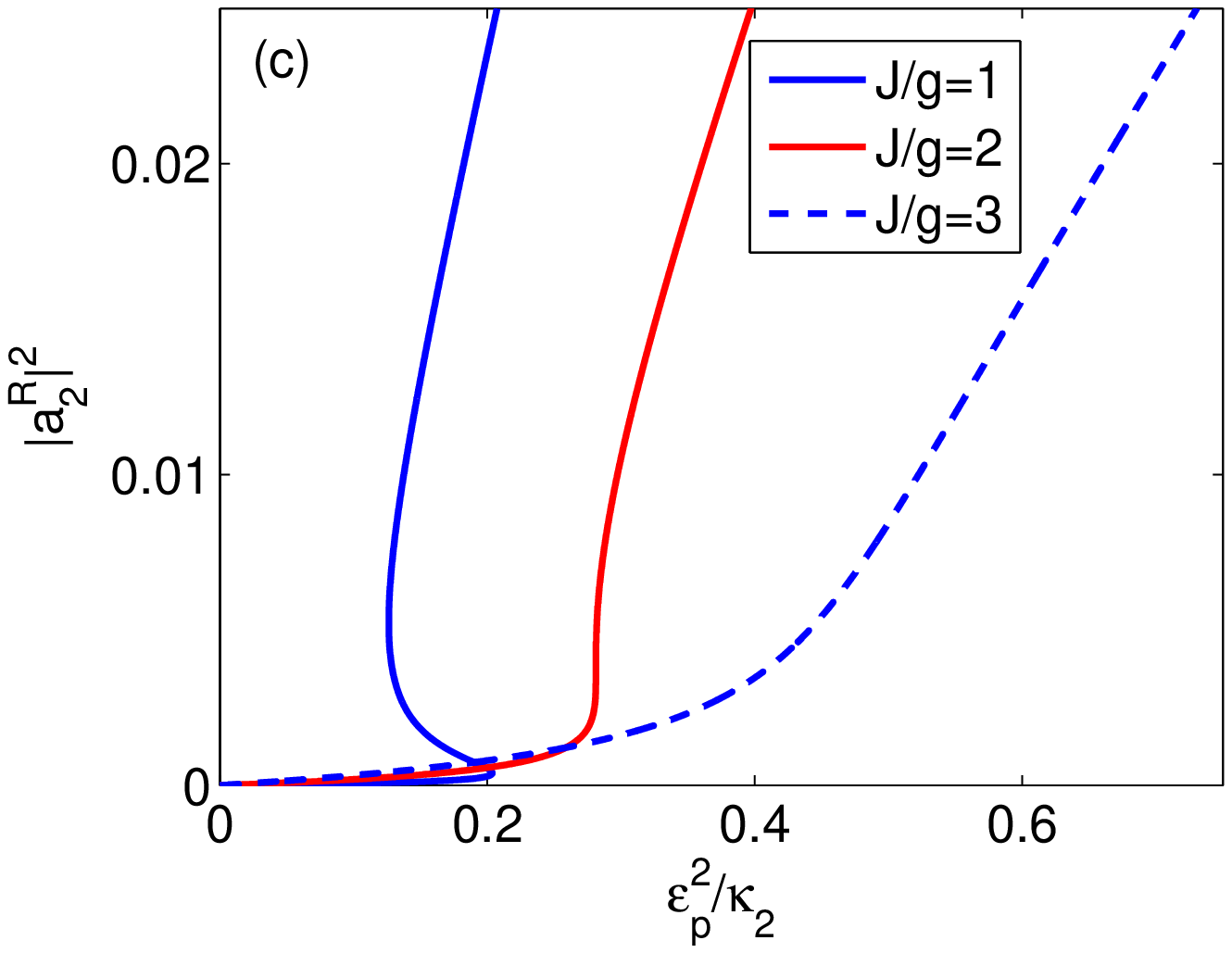}
\includegraphics[width=0.4\textwidth]{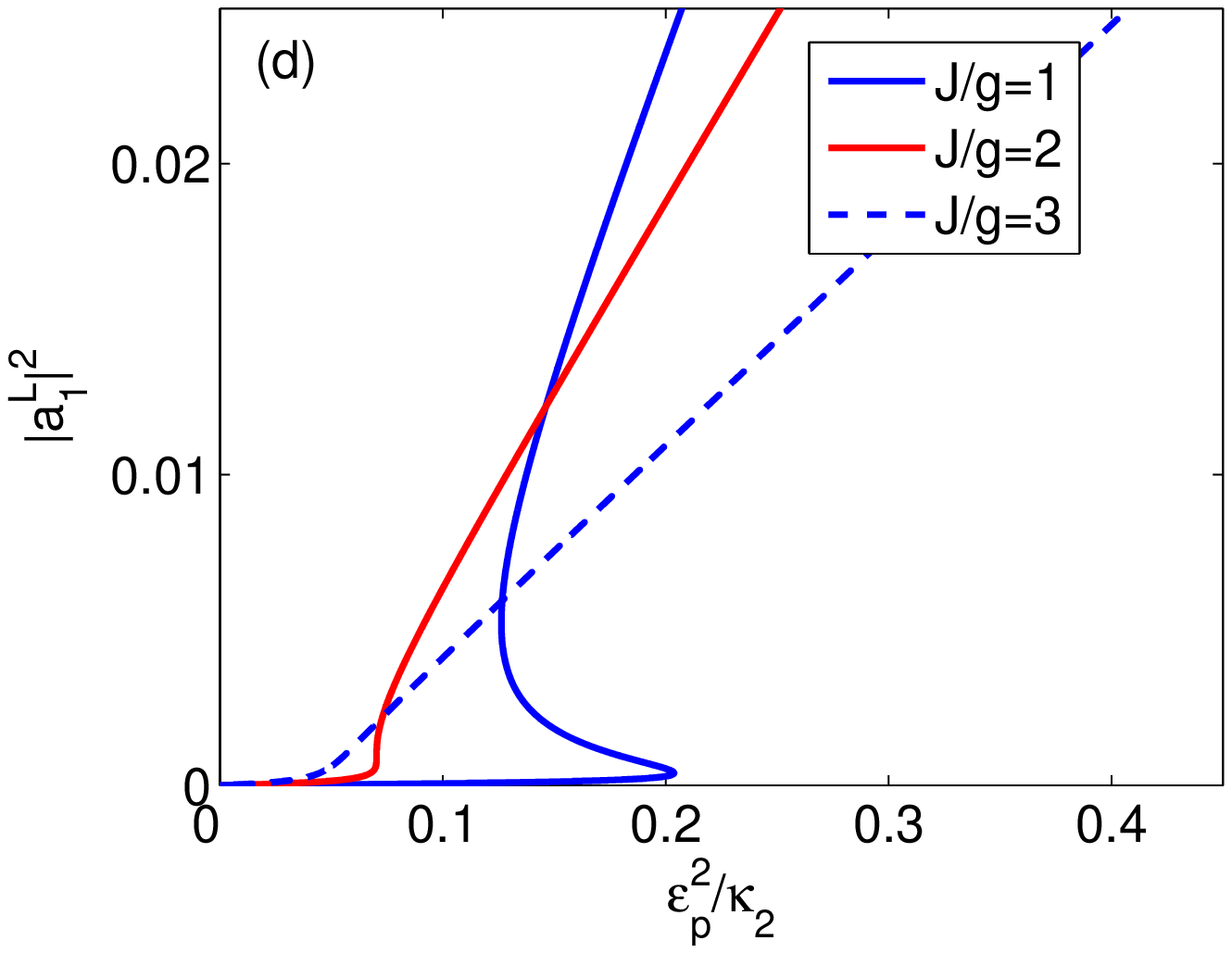}
\caption{\label{fig6} (Color online) For passive-passive cavity system, the output field $|a_2^R|^2, |a_1^L|^2$
as a function of the input field $\varepsilon_p^2/\kappa_2$ with (a), (b) J=4$\kappa_2$ and (c), (d) g=2$\kappa_2$.
The other system parameters are chosen as $\gamma=0.1\kappa_2$, $\Delta_1=\Delta_2=0$, $\kappa_1=\kappa_2$, and $\kappa_e=3\kappa_2$,
respectively.}
\end{figure}

Based on the Hamiltonian (1), the Heisenberg-Langevin equations, which describe the evolution of the optical composite system,
can be obtained.
We focus on the the mean response of the optical system, then the operators are reduced to their expectation values, and
the Heisenberg-Langevin equations lead to two groups of nonlinear evolution equations. In this work, our goal is to consider the
effect of the input field with different directions (i.e., forward and backward incidence) on the output field.
With the use of the mean-field assuption $\langle\hat{m}\hat{n}\rangle=\langle\hat{m}\rangle\langle\hat{n}\rangle$ \cite{j10},
two groups of nonlinear evolution equations of the hybrid system with the forward and backward incidence are
given as
\begin{eqnarray}\label{D3}
\dot{\alpha}_{L(R)}=M_{L(R)}\alpha_{L(R)}+\xi_{L(R)},
\end{eqnarray}
where the subscript $L$ and $R$ means the coupled microcavities is backward
or forward incidence, respectively. $\alpha_{L(R)}=(a_1^{L(R)},a_2^{L(R)},\sigma_z^{L(R)},\sigma_{ge}^{L(R)})^T$ with
$(a_1^{L(R)},a_2^{L(R)},\sigma_z^{L(R)},\sigma_{ge}^{L(R)})$ are the mean values of the corresponding operators of the hybrid system.
$\xi_L=(0,\sqrt{\kappa_e}\varepsilon_p,-\gamma/2,0)^T$, $\xi_R=(\sqrt{\kappa_e}\varepsilon_p,0,-\gamma/2,0)^T$,
\begin{equation}M_{L(R)}=\left(\begin{array}{cccc}
  x_1 & -iJ & 0 & -g \\
   -iJ & x_2 & 0 & 0 \\
    g\sigma_{ge}^{*L(R)} & 0 & -\gamma & ga_1^{*L(R)} \\
    -2g\sigma_z^{L(R)} & 0 & 0 & x_3\\
 \end{array}\right),
\end{equation}
where $x_1=-(i\Delta_1+\kappa_1/2+\kappa_e/2)$, $x_2=-(i\Delta_1+\kappa_2/2+\kappa_e/2)$, and
$x_3=-i(\Delta_1+\Delta_2)-\gamma/2$ with $\gamma$ the spontaneous emission decay rate and
$\kappa_j$ the cavity intrinsic decay rate. The evolution of the optical system varies with the direction of the input
beam. The phenomenon of direction-dependent evolution of the optical system can be displayed in the output
field, which can be obtained through the following input-output relation \cite{j42,j43}
\begin{eqnarray}
S_{out}^{R}=\sqrt{\kappa_e}a_2^R,\\
S_{out}^{L}=\sqrt{\kappa_e}a_1^L.
\end{eqnarray}

\section{numerical results and discussion about non-reciprocity light propagation}

In the present scheme, the strong nonlinearity of the system plays a significant role in enhancing
the ability of nonreciprocity light propagation. And it is induced by the interaction
between the optical cavity and quantum emitter. In the above nonlinear evolution equations (i.e., Eqs. (2)),
which describe the evolution of the hybrid optical system,
we pay attention to the nonlinear terms $ga_1^*\sigma_{ge}$, $ga_1\sigma_{ge}^*$ and $-2ga_1\sigma_{z}$.
With the weak-excitation approximation, these nonlinear terms are always discarded and the optical system will evolute
in linear regime. We value the nonlinear terms which lead to the
optical bistable state, as shown in Ref \cite{li1}. Then with the parameters in the bistable region, the considered optical system may have two different metastable
output values. Once the disturbance and perturbation of the system parameters become strong, the output field
will switch between the different metastale values. The uncertainty of the output field will be detrimental
to the nonreciprocity light propagation.
\begin{figure}[htb]
\includegraphics[width=0.4\textwidth]{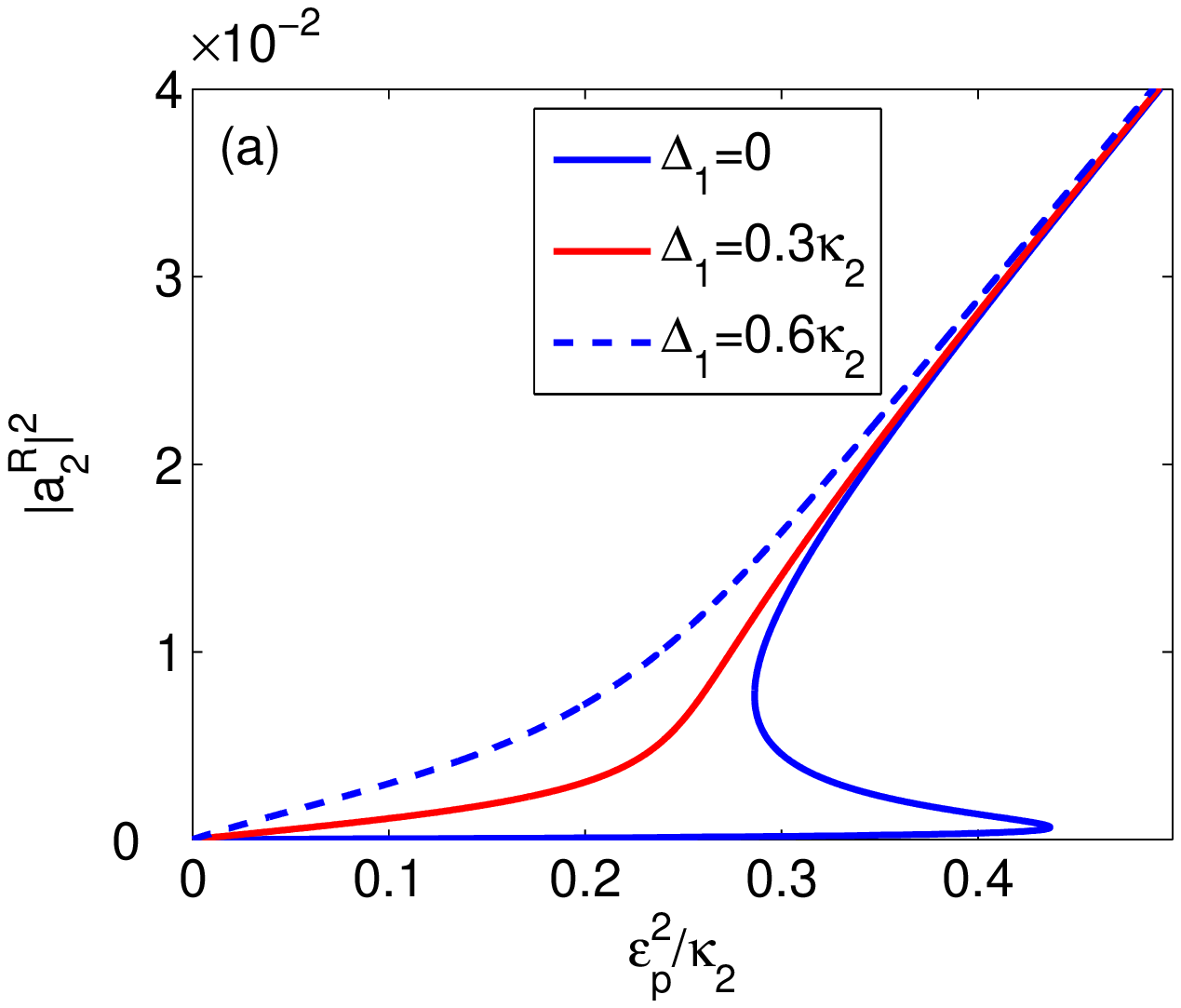}
\includegraphics[width=0.4\textwidth]{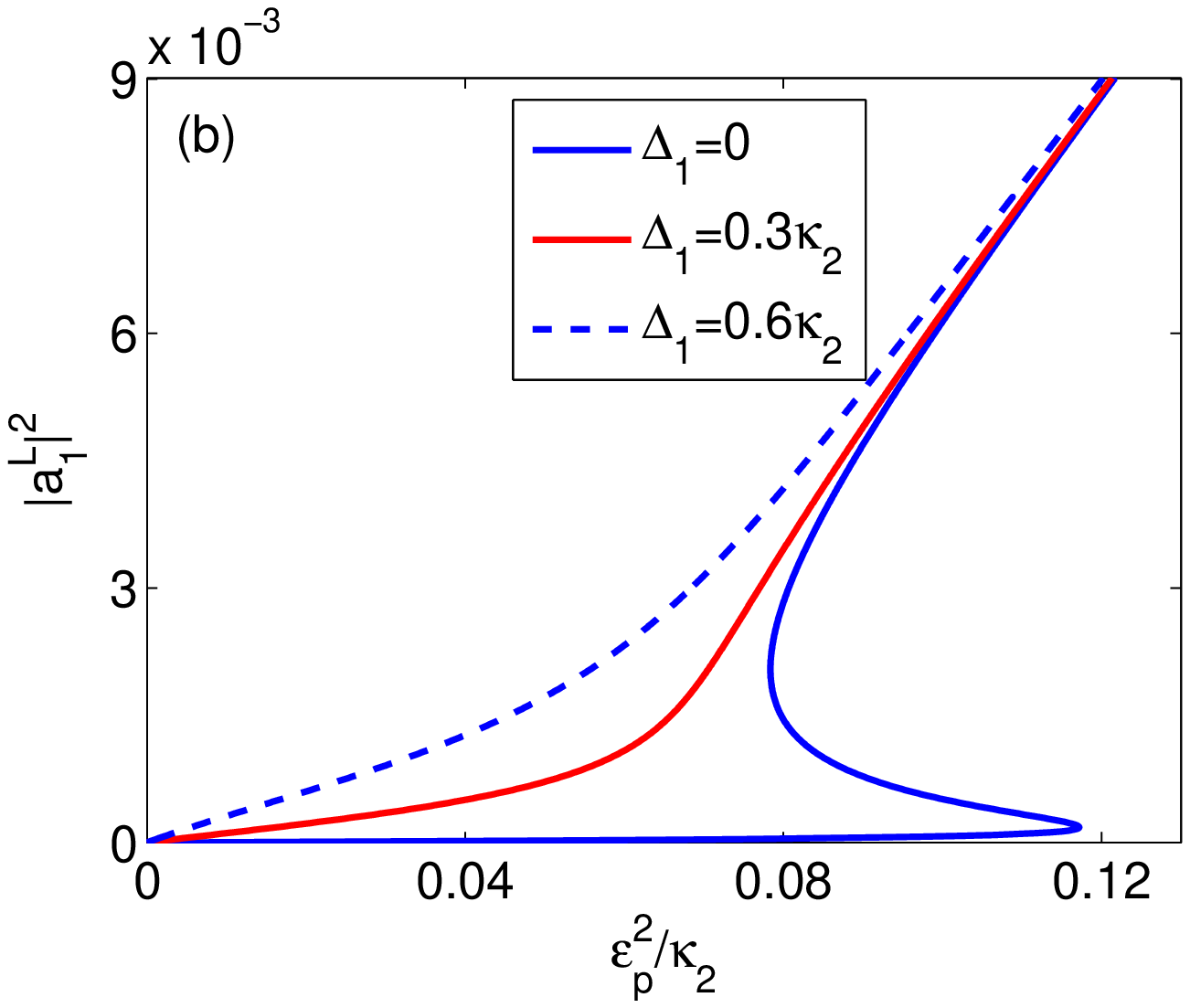}
\includegraphics[width=0.4\textwidth]{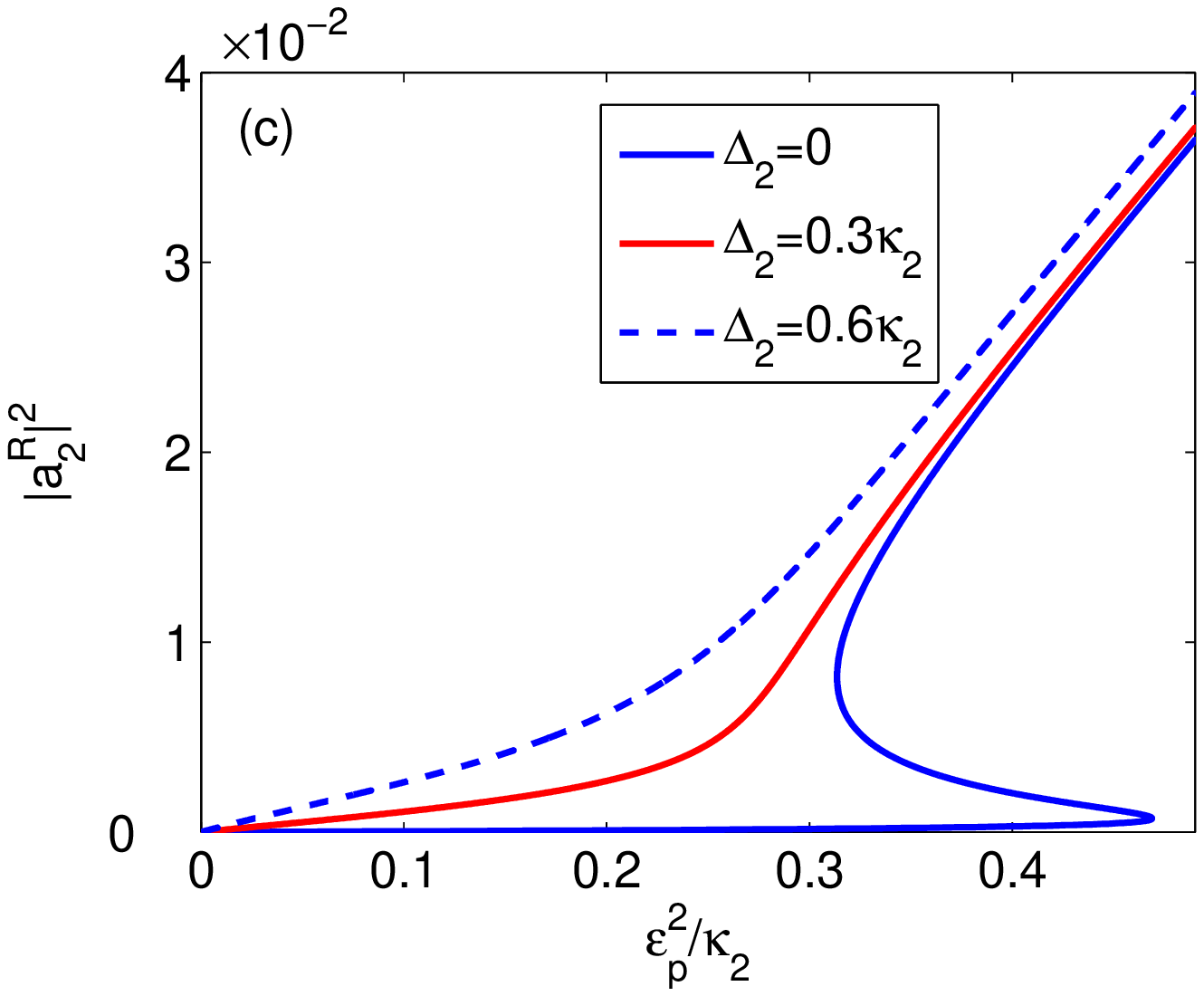}
\includegraphics[width=0.4\textwidth]{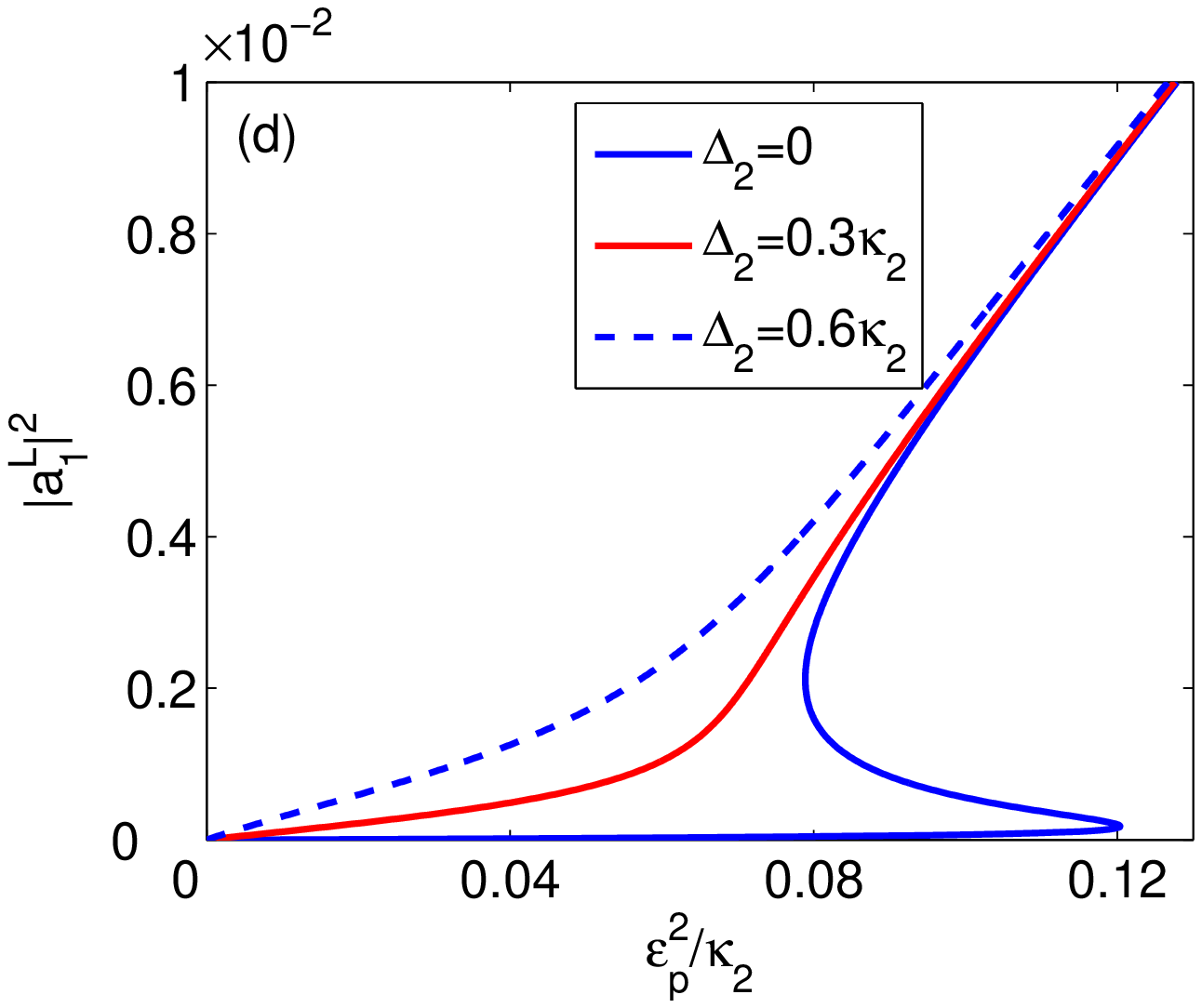}
\caption{\label{fig6} (Color online) The output field (a), (c) $|a_2^R|^2$, (b), (d) $|a_1^L|^2$
as a function of the input field $\varepsilon_p^2/\kappa_2$ with J=4$\kappa_2$ and g=2$\kappa_2$. The other system parameters
are the same as in Fig. 2.}
\end{figure}
\begin{figure}[htb]
\includegraphics[width=0.45\textwidth,height=0.25\textheight]{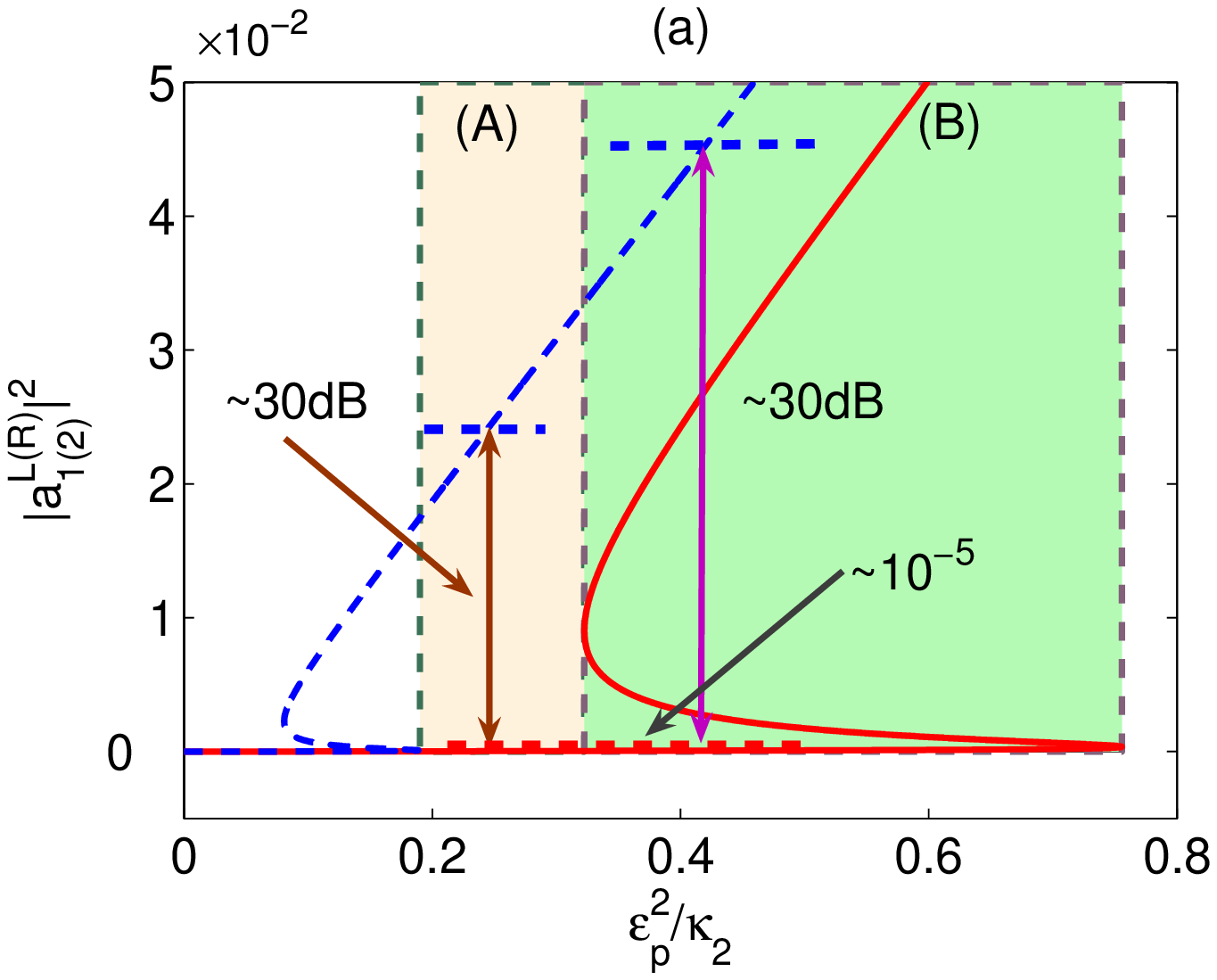}
\includegraphics[width=0.45\textwidth,height=0.25\textheight]{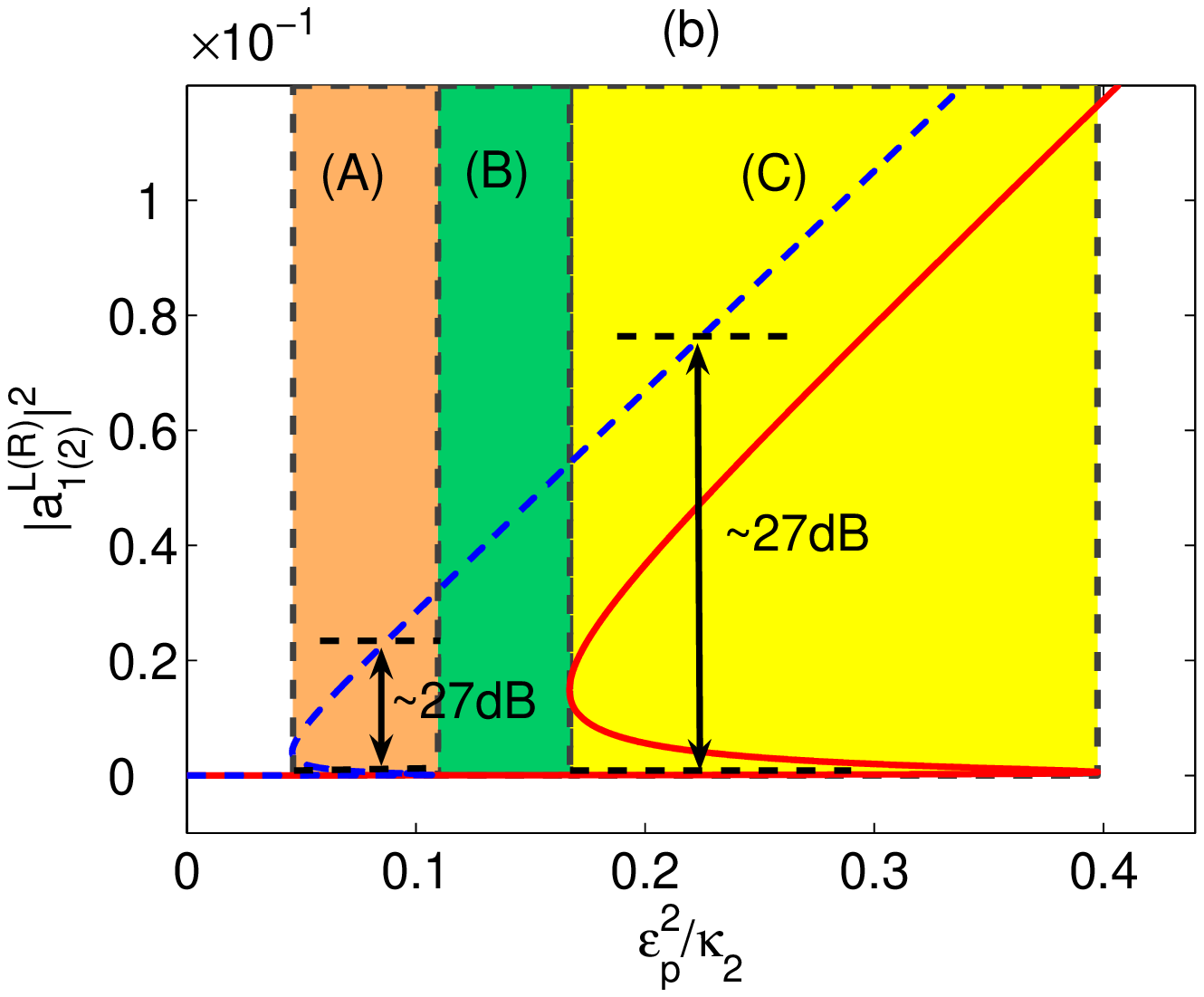}
\caption{\label{fig6} (Color online) The output field $|a_2^R|^2$ (red solid line) and $|a_1^L|^2$ (blue dashed line) vary the input field
 $\varepsilon_p^2/\kappa_2$ with (a) $\gamma=0.1\kappa_2$, $g=4\kappa_2$, $\Delta_1=\Delta_2=0$, $\kappa_1=\kappa_2$, $\kappa_e=3\kappa_2$, and $J=4\kappa_2$ for the passive-passive cavity system;
  (b) $\gamma=0.1\kappa_2$, $g=3\kappa_2$, $\Delta_1=\Delta_2=0$, $\kappa_1=-7.4\kappa_2$,
 $\kappa_e=3.2\kappa_2$, and $J=4\kappa_2$ for the unbroken $\textrm{PT}$ phase (i.e., active-passive cavity system).}
\end{figure}

In order to avoid the negative effect of the inconclusive output field on the nonreciprocal light propagation, we begin to carefully
consider the optical bistable behavior, i.e., how does the output field depend on the system parameters.
By seeking the numerical steady-state solution of Eqs .(2), we show the the steady-state output field intensity $P_{out}^R\propto|a_2^R|^{2}$ and $P_{out}^L\propto|a_1^L|^{2}$ versus the input filed intensity $P_{in}^{L(R)}\propto\varepsilon_p^2$ under various parametric conditions
in Fig. 2. The influence of the cavity-quantum emitter coupling strength $g$ on the behavior of the optical bistability
is shown in Fig. 2(a) and Fig. 2(b).
The bistable threshold increases gradually as the cavity-quantum emitter coupling strength $g$ increases. More importantly, the area
of the hysteresis loop becomes broader as the coupling strength $g$ increases from $g=2\kappa_2$
to $g=7\kappa_2$. Conversely, the optical bistable regions will disappear when the cavity-quantum emitter coupling strength is small enough.
This is because that the optical bistability is caused by the nonlinear terms $ga_1^*\sigma_{ge}$,
$ga_1\sigma_{ge}^*$ and $-2ga_1\sigma_{z}$ in Eqs. (2), and at a large extent, the increasing coupling strength $g$ between
the cavity and the quantum emitter can enhance the nonlinearity of the optical system.
On the other hand, the cavity-cavity coupling strength $J$ has also an important influence on the optical bistability.
There is a competition between the cavity-quantum emitter coupling strength $g$ and cavity-cavity coupling strength $J$ in
the input field of
the optical system as shown in Fig. 2(c) and (d). When the cavity-quantum emitter coupling strength and cavity-cavity coupling strength
are equal (i.e., $g=J$), there is a remarkable optical bistable region,  which quickly becomes narrow as the
ratio of the cavity-cavity coupling strength to cavity-quantum emitter coupling strength
 increases. Once the cavity-cavity coupling has an overwhelming advantage against the cavity-quantum emitter coupling
(i.e., $J\geq 3g$), the nonlinearity of the system can be neglected and the optical bistable area will disappear completely.
Besides the cavity-cavity coupling strength $J$, the frequency detunings $\Delta_j(j=1,2)$ also affect the cavity-quantum emitter
interaction, which induces the nonlinearity of the considered system. The increasing frequency detuning will weaken the
cavity-quantum emitter coupling and makes the optical bistable area small. As shown in Fig. 3, the optical bistable
area gradually becomes narrow until disappears as the frequency detuning increases.
\begin{figure}[htb]
\includegraphics[width=0.45\textwidth,height=0.2\textheight]{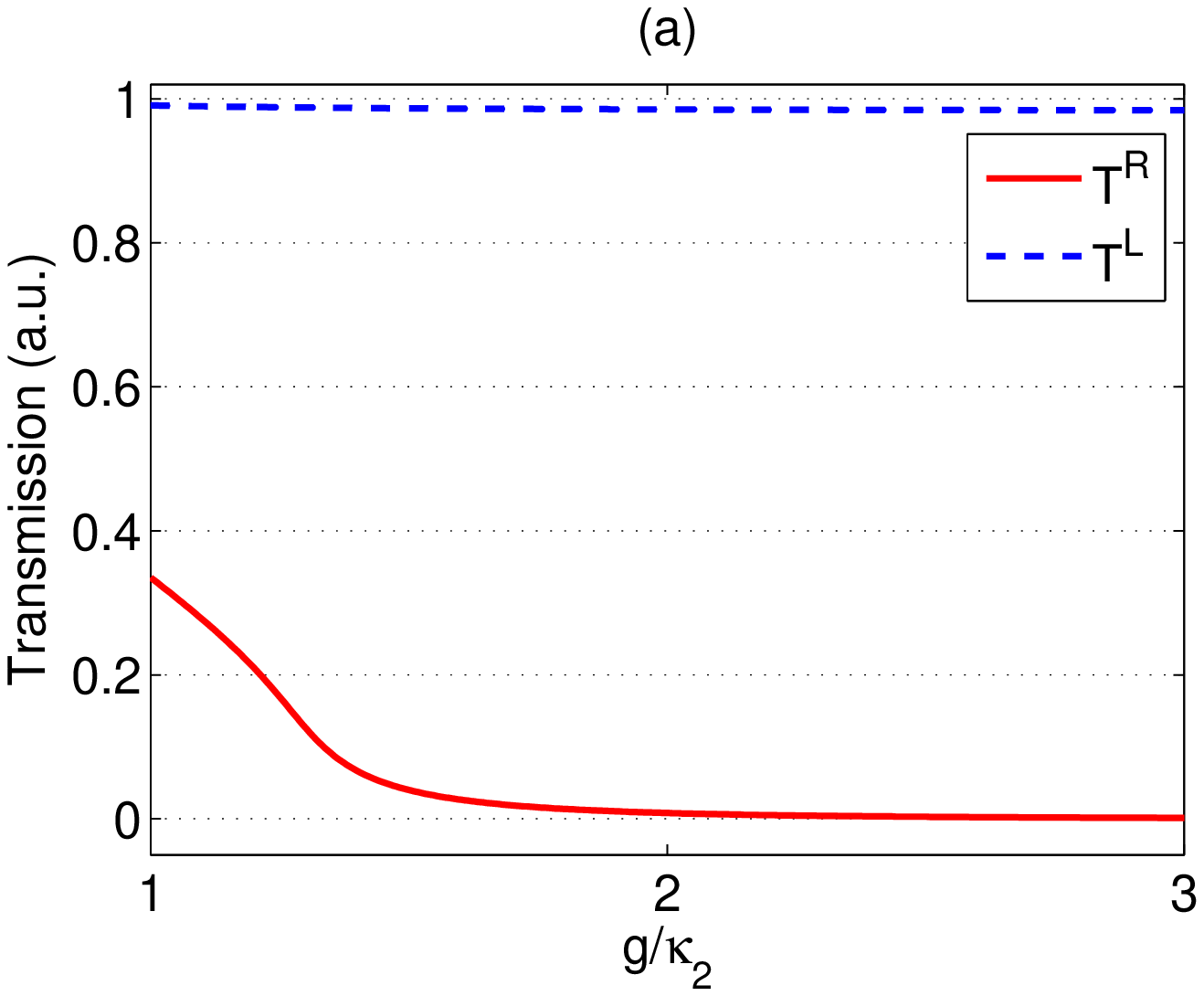}
\includegraphics[width=0.45\textwidth,height=0.2\textheight]{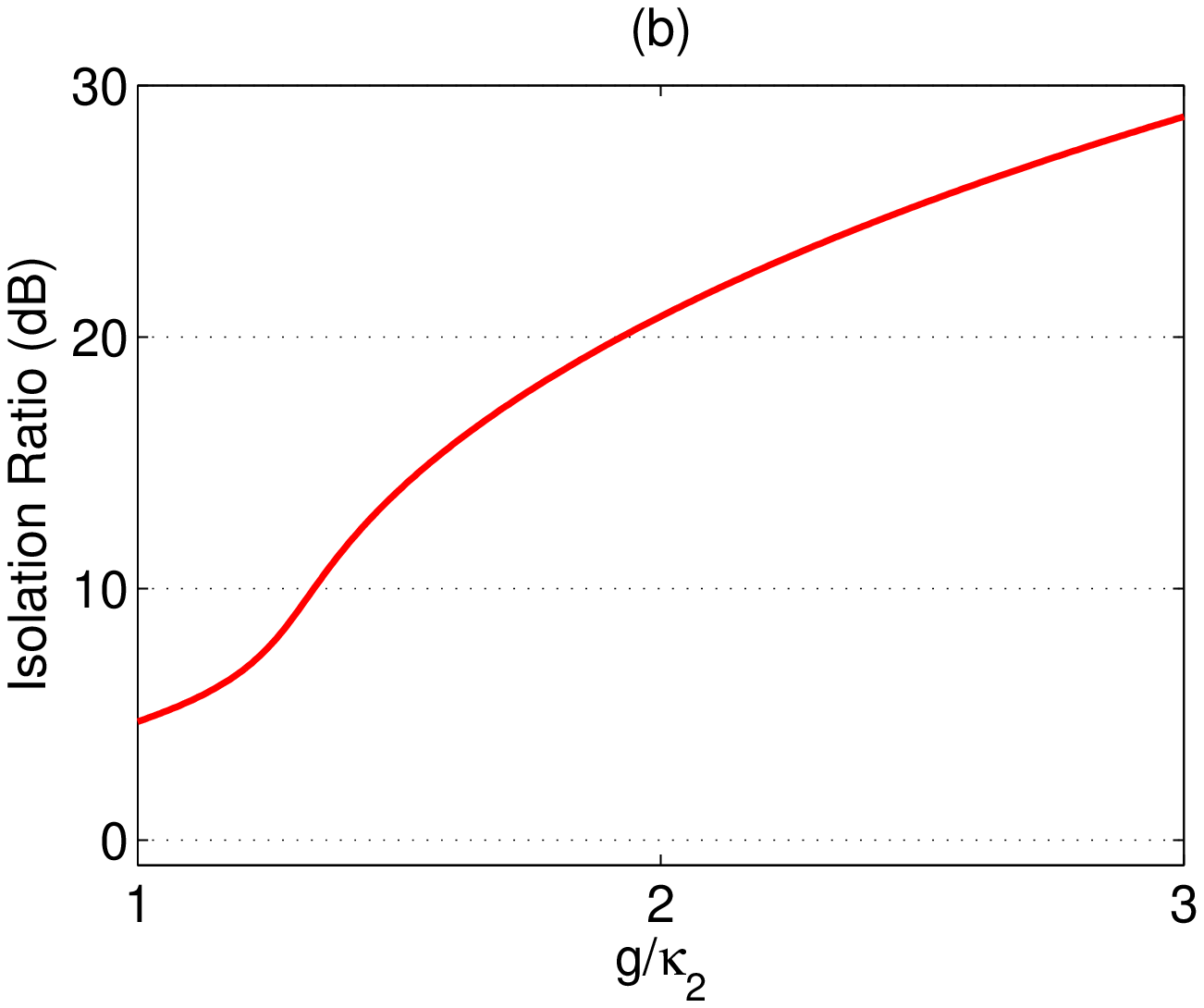}
\includegraphics[width=0.45\textwidth,height=0.2\textheight]{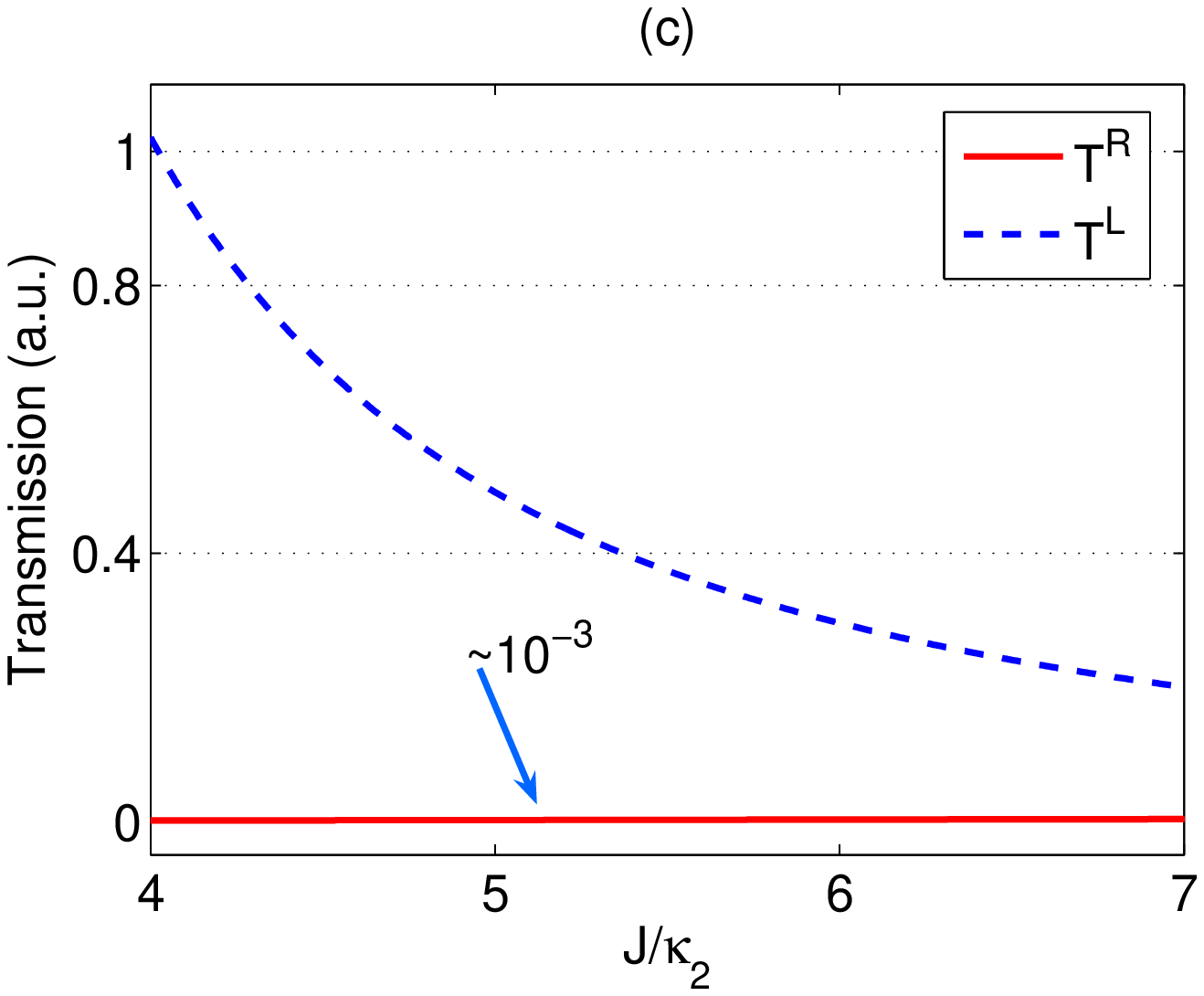}
\includegraphics[width=0.45\textwidth,height=0.2\textheight]{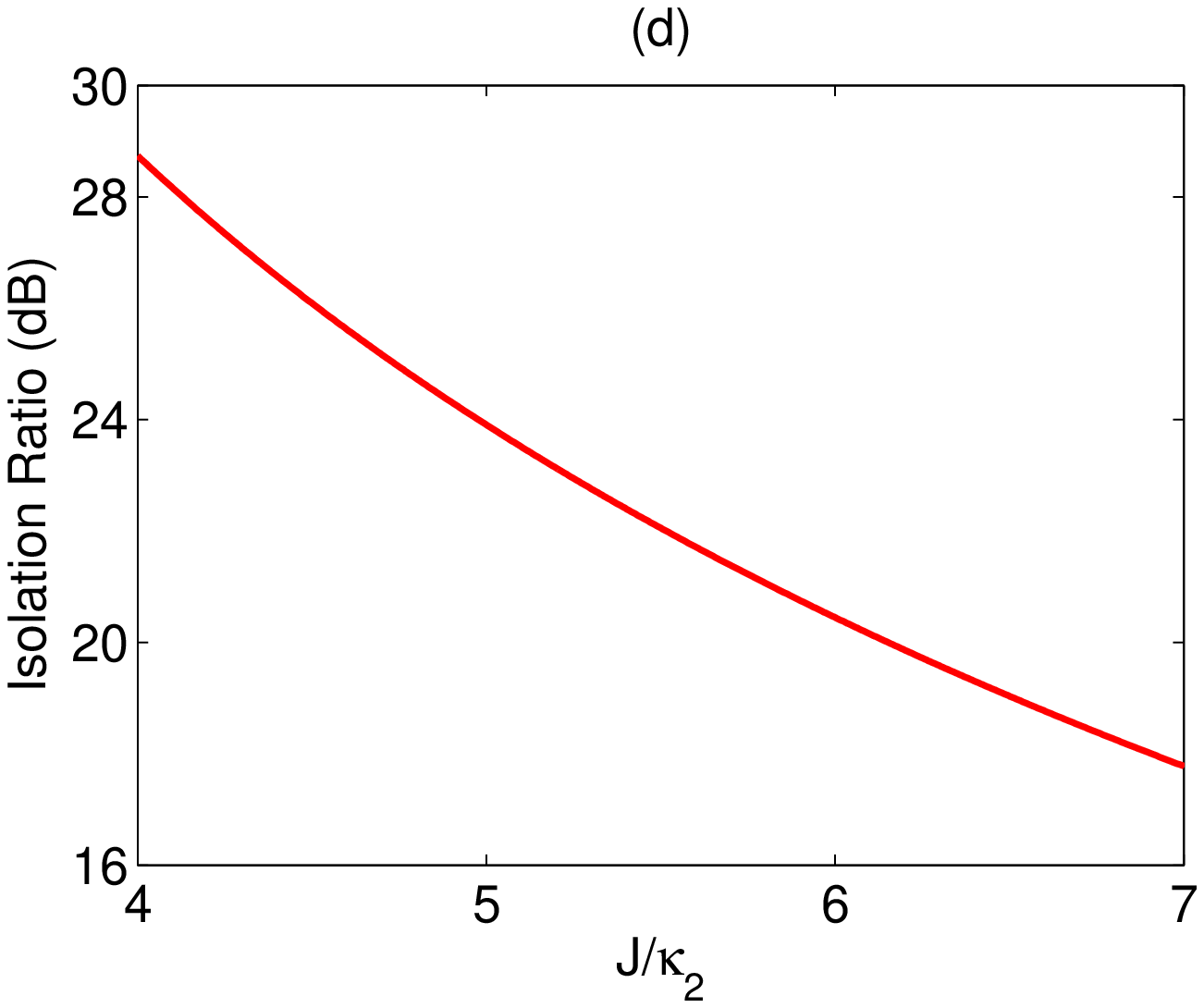}
\includegraphics[width=0.45\textwidth,height=0.2\textheight]{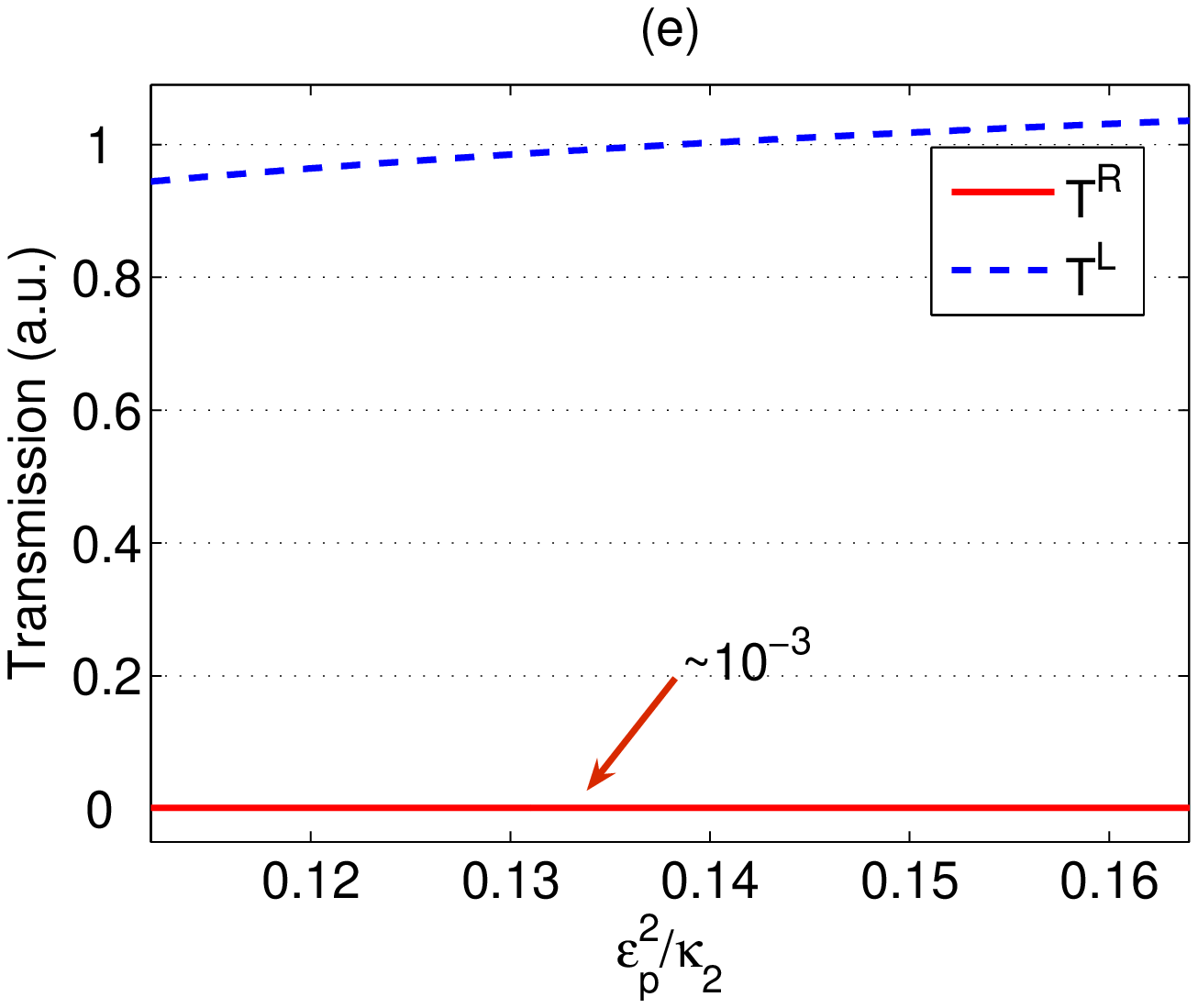}
\includegraphics[width=0.45\textwidth,height=0.2\textheight]{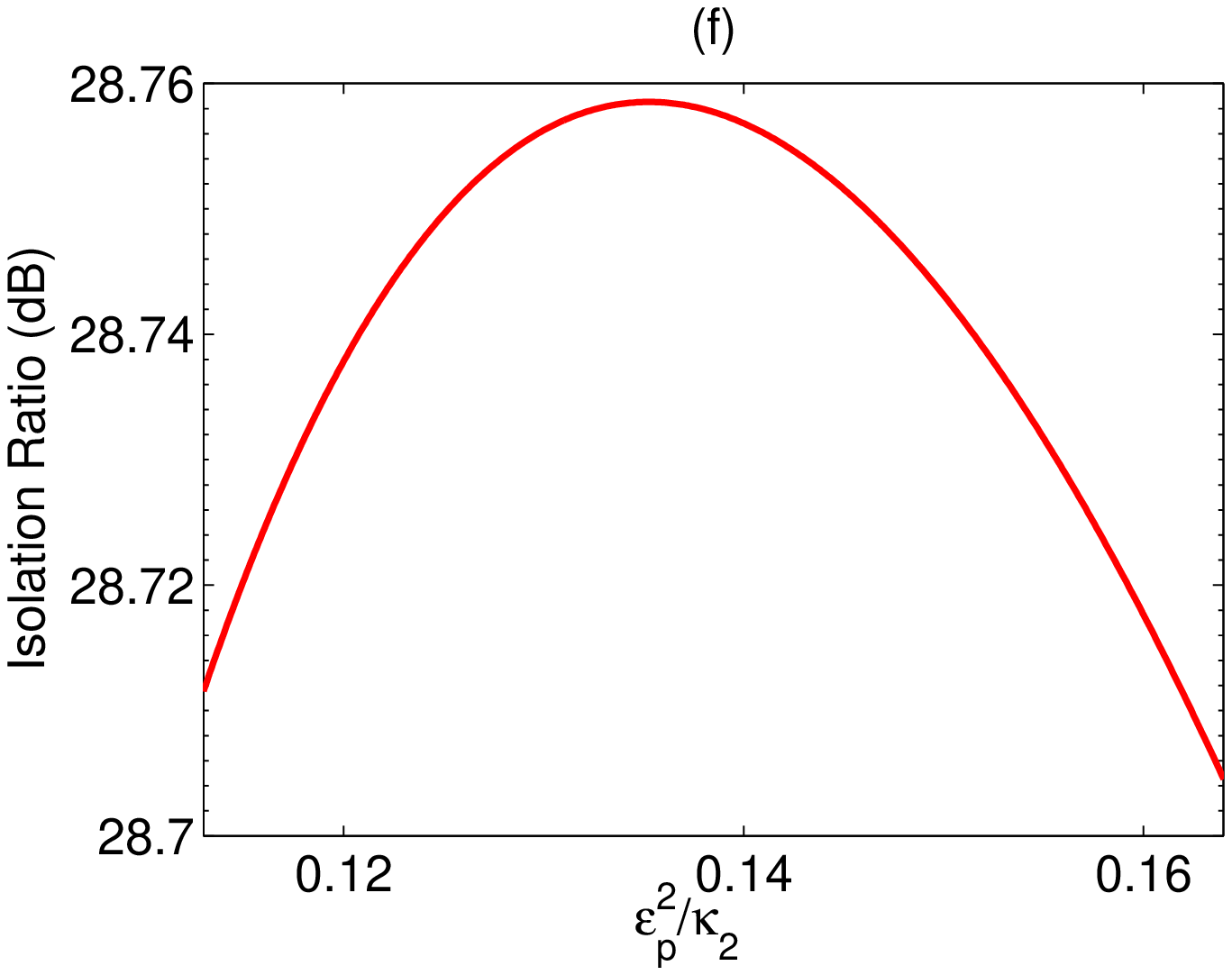}
\caption{\label{fig6} (Color online) For the resonant case, the transmission coefficient $T^R$ (red solid line) and $T^L$ (blue dashed line),
and the isolation ratio vary under the different values of (a),(b) the cavity-quantum emitter coupling strength $g/\kappa_2$ with $J=4\kappa_2$, and $\varepsilon_p=0.36\sqrt{\kappa_2}$;
(c), (d) the cavity-cavity coupling strength $J/\kappa_2$
with $\varepsilon_p=0.39\sqrt{\kappa_2}$, and $g=3\kappa_2$;
(e), (f) the input field $\varepsilon_p^2/\kappa_2$ with $g=3\kappa_2$, and $J=4\kappa_2$.
The other system parameters are chosen as $\gamma=0.1\kappa_2$, $\kappa_1=-7.4\kappa_2$, and $\kappa_e=3.2\kappa_2$, respectively.}
\end{figure}

Based on the above analysis, we begin to explore the relationship between the output and input intensity for the
forward and backward propagation cases. The transmission coefficients
are defined as
\begin{equation}
T^{L(R)}=|\frac{S_{out}^{L(R)}}{S_{in}}|^2,
\end{equation}
and the isolation ratio is
\begin{equation}
Isolation\  Ratio (dB)=10\times \log_{10}{\frac{T^L}{T^R}},
\end{equation}
which quantifies the isolation performance of the system.
Fig. 4(a) and Fig. 4(b) correspond to the passive-passive cavities coupling and active-passive
cavity coupling, respectively. In the color-code region (B) of Fig. 4(a), the maximum isolation ratio is approximately $30$dB
when the output field intensity for the forward propagation stays at the lower launch. However, in the color-code region (A) of Fig. 4(a),
both the forward and backward propagation have only one stable output value. Thus, if we choose the system parameters in the
color-code region (A) of Fig. 4(a), the shortcoming of the color-code region (B)
of Fig. 4(a) (i.e., the uncertain
output field intensity) caused by the optical bistability, can be overcomed, and we can also obtain the high isolation ratio.
The physical mechanism underlying the nonreciprocal light transport is rooted in
the cavity-quantum emitter interaction inducing nonlinearity, which leads to localization-induced
dynamical-intensity accumulation in the first cavity. The asymmetrical coupling breaks the time-reversal symmetry, which
make the nonreciprocal light transport feasible, i.e., the light transport from the cavity $2$ to the cavity $1$
is allowed and the light transport of the opposite direction is blocked.
For the passive-passive cavity system, due to the decay rate of the cavities and emitter,
 the low transmissivity (i.e.,weak output field intensity, about $30\%$ of the input field
intensity) is another obstacle  to realize the nonreciprocal light transport.
\begin{figure}[htb]
\includegraphics[width=0.45\textwidth,height=0.25\textheight]{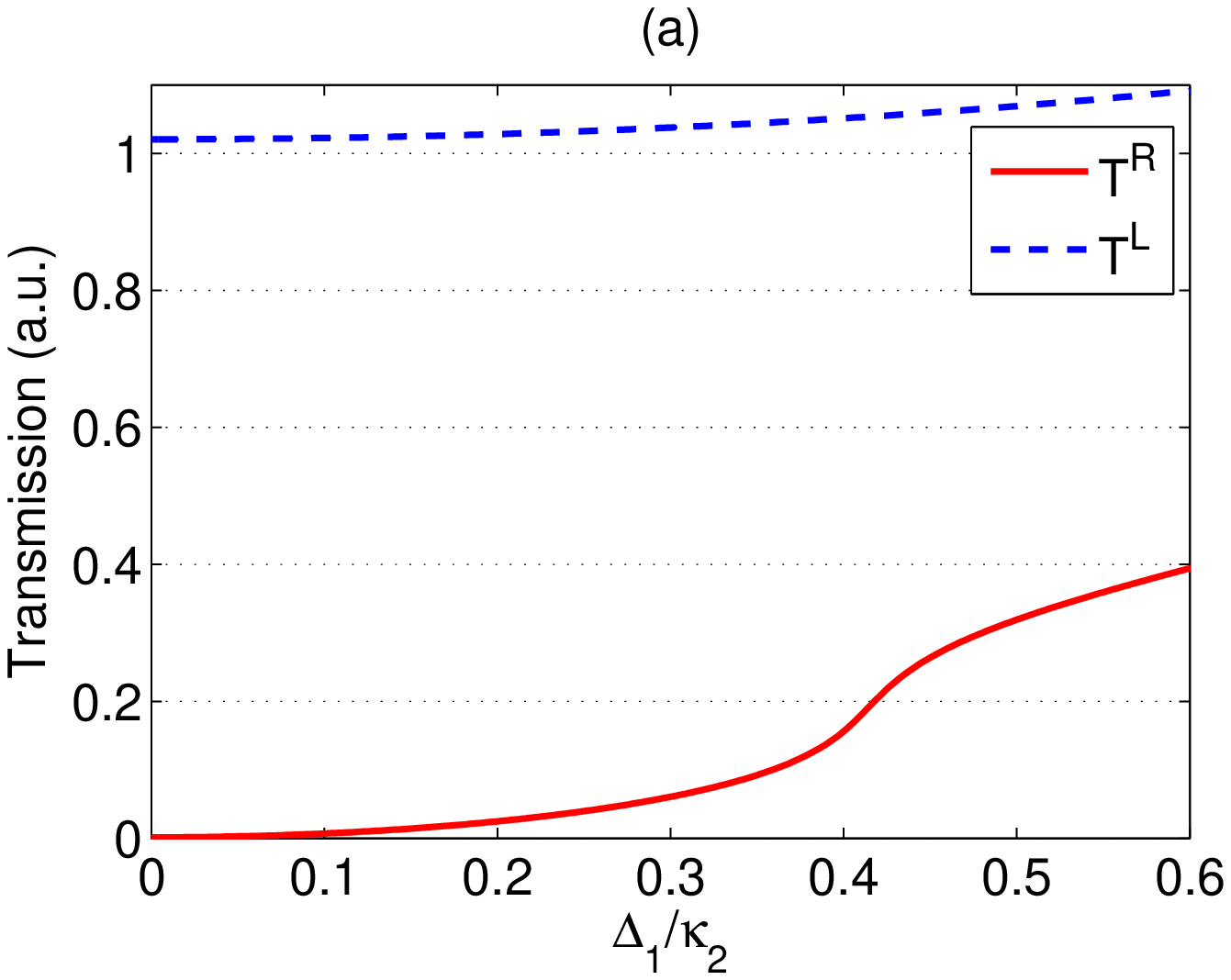}
\includegraphics[width=0.45\textwidth,height=0.25\textheight]{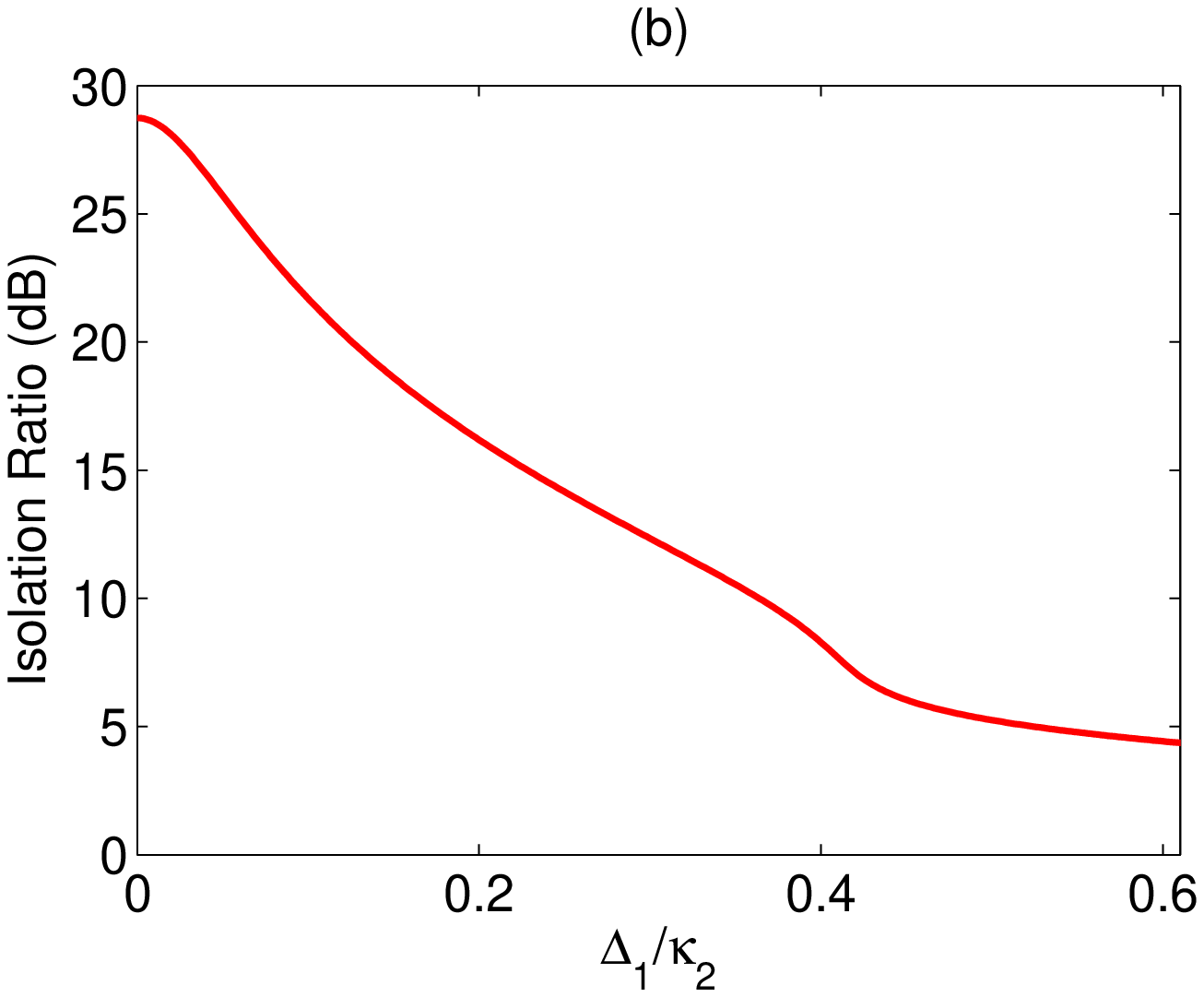}
\includegraphics[width=0.45\textwidth,height=0.25\textheight]{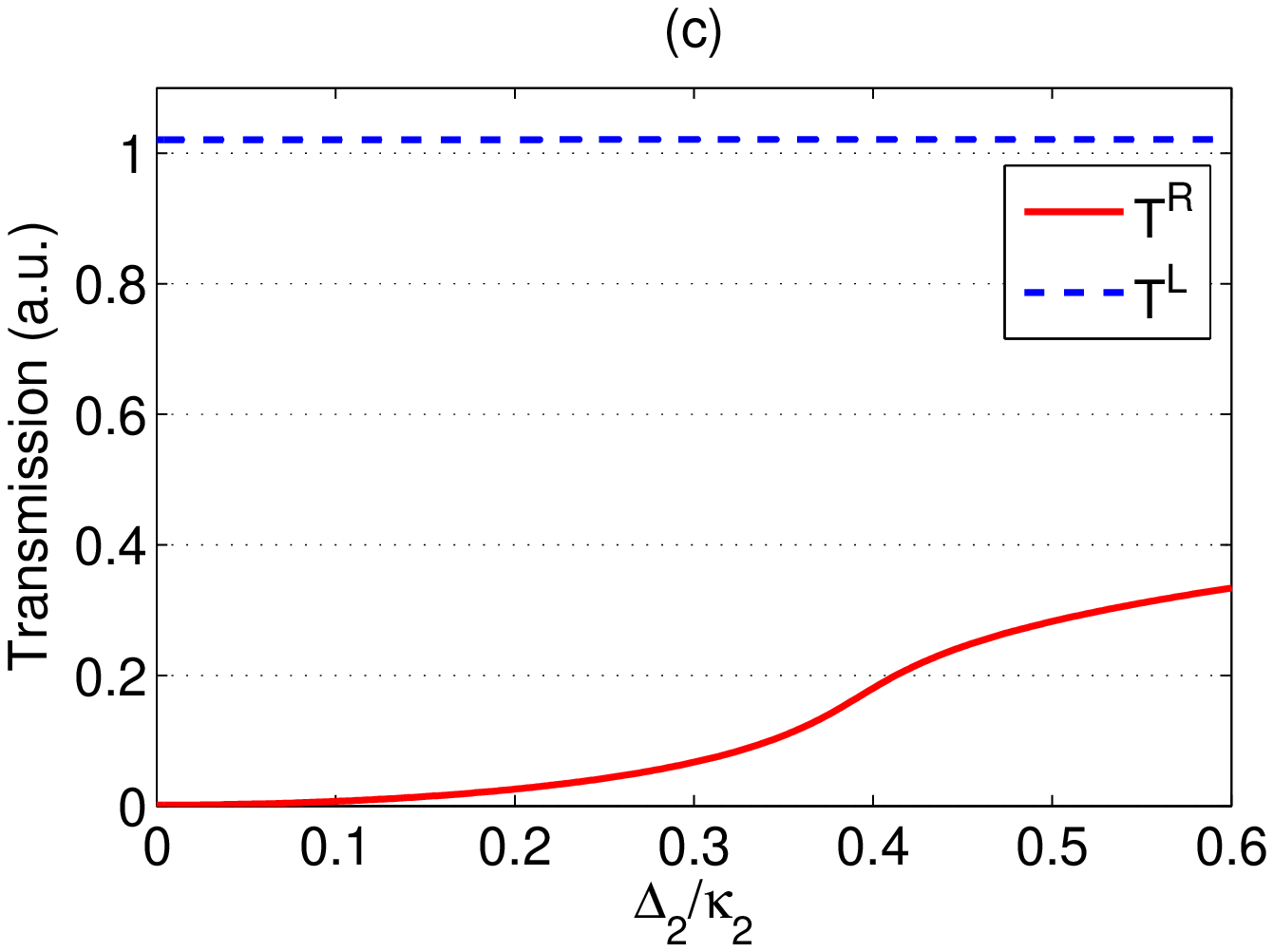}
\includegraphics[width=0.45\textwidth,height=0.25\textheight]{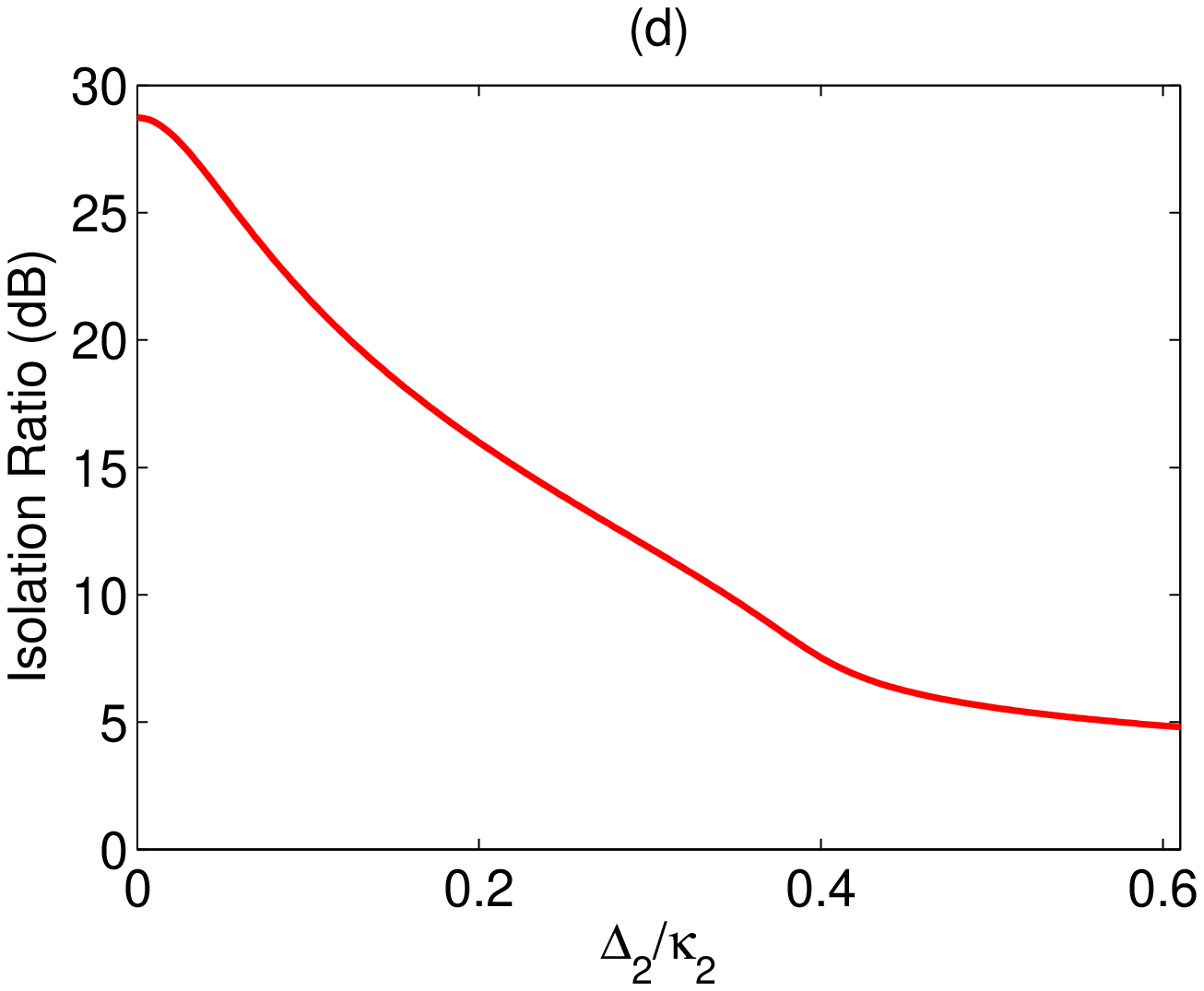}
\caption{\label{fig6} (Color online) For the off-resonant case, the transmission coefficient
$T^L$ (blue dashed line) and $T^R$ (red solid line), and isolation ratio vary with (a), (b) frequency
detuning $\Delta_1/\kappa_2$ for $\Delta_2=0$; with (c), (d) frequency detuning $\Delta_2/\kappa_2$ for $\Delta_1=0$.
The other system parameters are chosen as $\gamma=0.1\kappa_2$, $\kappa_1=-7.4\kappa_2$, $\kappa_e=3.2\kappa_2$, $J=4\kappa_2$,
 $\varepsilon_p=0.36\sqrt{\kappa_2}$, and $g=3\kappa_2$, respectively.}
\end{figure}

In order to enhance the transmissivity, we can choose the \textit{PT}-symmetric system
(i.e., the active-passive cavity system) as shown in Fig. 4(b).
The gain of the first cavity greatly improves the nonlinearity and promotes the field localization in the
first cavity. It allows the light propagation from the passive cavity to the active cavity and prevents
the propagation in the opposite direction \cite{24nat,25nat}.
Thus, the gain-loss balance of the \textit{PT}-symmetric system makes the non-lossy unidirectional light transport achievable.
As shown in all the color-code areas of Fig. 4(b),
we can obtain the nonreciprocal light transport
with over $99\%$ high transmissivity and about $27$ dB isolation ratio. Among of the color-code areas of Fig. 4(b),
the area (B) is more suitable for the idea unidirectional light transport
without the disturbance of the uncertain output field intensity caused by optical bistability because the output field intensity
for the forward and backward propagations have only one stable value.

With respect to the color-code area (B) of Fig. 4(B), we begin to analyze the effect of the experimental parameters deviation on the unidirectional light transport in detail. Other parameters are fixed, and the non-lossy unidirectional light propagation from passive cavity to
 active cavity is not sensitive to the cavity-quantum emitter coupling. When the cavity-quantum emitter coupling $g$
 is close to zero, the linear system allows the non-lossy light propagation in both directions with the help of
gain of the first cavity, but the isolation ratio declines sharply as shown in Fig. 5(a) and Fig. 5(b). For the growing
coupling strength $g$, the nonlinearity of the system will also tend to increase and will be greatly enhanced by the gain of the first cavity.
The large nonlinearity is only in the first cavity (i.e., the active cavity) and promotes the field localization in
the active cavity and breaks the time-reversal symmetry of the considered optical system. Thus, the non-lossy light propagation in the backward direction
is almost unaffected and the opposite direction propagation is blocked completely. When the cavity-quantum emitter coupling
strength $g$ approaches $3\kappa_2$, we can obtain the isolation ratio about $27$ dB. Fig. 5(c) and Fig. 5(d) show the influence
of cavity-cavity coupling $J$ on the unidirectional light propagation. The cavity-cavity coupling $J$ and the cavity-quantum emitter
coupling $g$ compete for the input field of the system. When the cavity-quantum emitter coupling $J$ is in the commanding position,
the nonlinearity of system will decrease. Thus, the light transmissivity in the backward direction reduces with the
decrease of cavity-cavity coupling, and the isolation ratio goes down as well.
From Fig. 5(c) and Fig. 5(d), we can see the effect of the change of the input field intensity on the unidirectional light
propagation can be neglected. In addition, the frequency detuning has different
influence from the cavity-cavity coupling on the present scheme. As the frequency detuning increases,
the nonlinearity of the system decreases. Fig. 6 shows the concrete effect of the frequency detuning on the
unidirectional light propagation.
\begin{figure}[htb]
\includegraphics[width=0.45\textwidth,height=0.25\textheight]{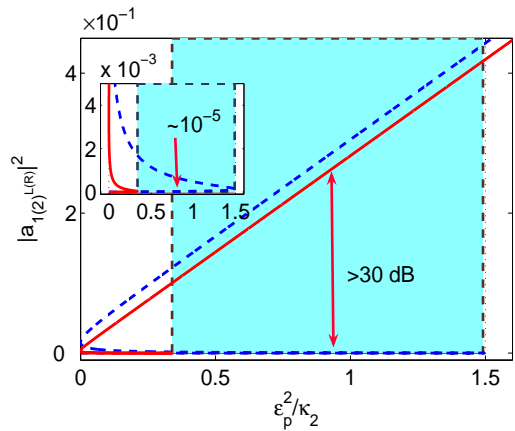}
\caption{\label{fig6} (Color online) The output field $|a_1^L|^2$ (blue dashed line) and $|a_2^R|^2$ (red solid line)
vary with the input field $\varepsilon_p^2/\kappa_2$ for the resonant case. The parameters are
$\gamma=0.1\kappa_2$, $g=3\kappa_2$, $\kappa_1=-7.4\kappa_2$, $\kappa_e=3.2\kappa_2$, and $J=\kappa_2$.}
\end{figure}

We begin to describe an interesting feature of the optical system, i.e., the allowed direction of the unidirectional
 light propagation can be reversed by adjusting the parameter $J$. On the above analysis, the light propagation in backward direction
is allowed and the opposite direction is blocked. When other parameters are fixed and the cavity-cavity coupling strength
is weaker than the cavity-quantum emitter coupling strength (for example, $J=\kappa_2$ and $g=3\kappa_2$), the nonlineraity
of the system is strengthened. As shown in Fig. 7, the optical bistable threshold decreases rapidly and approaches to zero
for both directions, and the
optical bistable area is clearly expending especially for the backward direction. As a result, the optical bistable area overlaps for the different directions.
In the color-code area of Fig. 4, we choose the lowest branch of the blue dash line which corresponds to the
light propagation from the active cavity to the passive cavity, and then we get the high efficiency unidirectional
light propagation with the isolation ratio about $30$ dB. In addition, when the input field intensity increases and exceeds
the corresponding maximum value in
the optical bistable region (for example, $\varepsilon_p^2>1.5\kappa_2$ in Fig. 7), the present system will turn to the bidirectional transport regime.

\section{Conclusion}
In summary, we have proposed a theory scheme for nonreciprocal light propagation with the
cavity-cavity coupling hybrid system. We analyze the effect of the system parameters on the optical bistable behavior,
and further study the nonreciprocal light propagation behavior in the passive-passive cavity system.
However, the interaction of the cavity-quantum emitter causes
 the weak nonlinearity, which is greatly enhanced by the cavity gain in the active-passive cavity system
 (i.e., \textrm{PT}-symmetric system).
 Thus, through balancing
the gain and loss, we can obtain the non-lossy and high isolation ratio nonreciprocal light propagation in the unbroken \textrm{PT} phase.
With appropriate
parameters, we can eliminate the risk of the uncertain output field intensity from optical bistability.
In addition, the direction of the nonreciprocal light propagation can be switched by changing
the cavity-cavity coupling strength $J$, which is sensitive to the distance between the cavities.

\section*{Acknowledgments}
The work is supported in part by National Basic Research
Program of China (no 2012CB922103) and by the National
Science Foundation (NSF) of China (grant nos 11374116) and the Fundamental Research Funds for the
Central Universities. The authors acknowledge
Professor Ying Wu for his enlightening suggestions and
anonymous referees for their insightful comments.

\end{document}